\documentclass[final,12pt]{elsarticle}
\usepackage{lineno,hyperref}
\usepackage[utf8]{inputenc}
\usepackage{graphicx} % include figures
\usepackage{amssymb} % AMS math symbol
\usepackage{url}     % handle url
\usepackage{bm}      % bold math
\usepackage{color}
\usepackage{lipsum}
\usepackage{hyperref}
\usepackage{subfig}
\usepackage{float}
\usepackage{amsmath}
\usepackage{booktabs}
\usepackage{bibunits}

\def\pmat{\ensuremath{\begin{pmatrix}}}

\def\CH{\ensuremath{{\mathcal{H}}}}

\def\be{\begin{eqnarray}}
\def\ee{\end{eqnarray}}
\def\bea{\begin{eqnarray}}
\def\eea{\end{eqnarray}}
\def\beas{\begin{eqnarray*}}
\def\eeas{\end{eqnarray*}}

\def\smat{\left[\begin{matrix}}
\def\emat{\end{matrix}\right]}

\def\nn{\nonumber}

\def\l{\left(}
\def\r{\right)}
\usepackage[symbol]{footmisc}

\newcounter{daggerfootnote}

\modulolinenumbers[5]
\defaultbibliography{ISOTRAP}
\defaultbibliographystyle{elsarticle-num}

\begin{document}

\author[HUJI,SARAF]
{
Yonatan Mishnayot
}

\author[HUJI]{Ayala Glick-Magid}
\author[HUJI]
{
Hitesh Rahangdale
}
\author[HUJI]
{
Guy Ron
\corref{mycorrespondingauthor}
}
\cortext[mycorrespondingauthor]{Corresponding author}
\ead{guy.ron2@mail.huji.ac.il}
\author[HUJI]
{
Doron Gazit
}
\author[LLNL]
{
Jason T. Harke
}
\author[WI]
{
Michael Hass\fnmark[1]
}
\fntext[1]{Deceased.}

\author[HUJI,ETH]
{
Ben Ohayon
}
\author[LLNL]
{
Aaron Gallant
}
\author[LLNL]
{
Nicholas D. Scielzo
}
\author[SARAF]
{
Sergey Vaintraub
}
\author[LLNL]
{
Richard O. Hughes
}
\author[SARAF]
{
Tsviki Hirsh
}
\author[Chalmers]
{
Christian Forss\'en
}
\author[Rez]
{
Daniel Gazda
}
\author[TRIUMF,UBC]
{
Peter Gysbers
}
\author[Barcelona]
{
Javier Men\'endez
}
\author[TRIUMF]
{
Petr Navr\'atil
}
\author[SARAF]
{
Leonid Weissman
}

\author[SARAF]
{
Arik Kreisel
}

\author[SARAF]
{
Boaz Kaizer
}

\author[SARAF]
{
Hodaya Dafna
}
\author[SARAF]
{
Maayan Buzaglo 
}

%\author[SARAF]{Soreq Nuclear Research Center, %Yavne, 81800\corref{mycorrespondingauthor}}

\address[HUJI]{The Racah Institute of Physics, The Hebrew University of Jerusalem, Givat Ram, Jerusalem, 9190401}
\address[SARAF]{Soreq Nuclear Research Center, Yavne, 8180000}
\address[LLNL]{Lawrence Livermore National Laboratory, Livermore, CA, USA}
\address[WI]{Department of Particle Physics, Weizamnn Institute of Science, Rehovot, Israel}

\address[ETH]{Institute for Particle Physics and Astrophysics, ETH Z\"urich, CH-8093 Z\"urich, Switzerland}
\address[Chalmers]{Department of Physics, Chalmers University of Technology, SE-412 96 G\"oteborg, Sweden}
\address[Rez]{Nuclear Physics Institute, 25068 \v{R}e\v{z}, Czech Republic}
\address[TRIUMF]{TRIUMF, 4004 Wesbrook Mall, Vancouver, British Columbia V6T 2A3, Canada}
\address[UBC]{Department of Physics and Astronomy, University of British Columbia, Vancouver, British Columbia, Canada}
\address[Barcelona]{Department of Quantum Physics and Astrophysics and Institute of Cosmos Sciences, University of Barcelona, 08028 Barcelona, Spain}

\title{Constraining new physics with a novel measurement of the $^{23}$Ne $\beta$-decay branching ratio}

\begin{abstract}
Measurements of the beta-neutrino correlation coefficient (a$_{\beta\nu}$) in nuclear beta decay, together with the Fierz interference term (b$_F$), provide a robust test for the existence of exotic interactions beyond the Standard Model of Particle Physics. The extraction of these quantities from the recoil ion spectra in $\beta$-decay requires accurate knowledge, decay branching ratios, and high-precision calculations of higher order nuclear effects. Here, we report on a new measurement of the $^{23}$Ne $\beta$-decay branching ratio, which allows a reanalysis of existing high-precision measurements. Together with new theoretical calculations of nuclear structure effects, augmented with robust theoretical uncertainty, this measurement improves on the current knowledge of a$_{\beta\nu}$ in $^{23}$Ne by an order of magnitude, and strongly constrains the Fierz term in beta decays, making this one 
of the first extractions to constrain both terms simultaneously. Together, these results place bounds on the existence of exotic tensor interactions and pave the way for new, even higher precision, experiments.
\end{abstract}

\flushbottom
\maketitle
\thispagestyle{empty}
\begin{bibunit}
The Standard Model of Particle Physics (SM) is currently the best description of particle interactions at our disposal. It predicts accurately processes as diverse as high-energy collisions in the Large Hadron Collider, to astrophysical reactions in the core of stars, which produce the elements of life. It is a well-known fact, however, that the SM is incomplete, since it does not fully describe 
observations such as dark matter, dark energy, the matter-antimatter imbalance,
and neutrino masses. Nor does it provide an {\it{ab initio}} explanation of a multitude of observations (such as, the absolute parity violation in the weak interaction). 

Experimental efforts at measuring beyond SM phenomena range from high-energy experiments (such as those performed at the LHC), to small lab-scale experiments, accessible on a table-top. At low energies, one of the most sensitive techniques for detecting Beyond SM (BSM) physics is through the  investigation of angular correlations in the $\beta$-decay of radioactive nuclei. The canonical $\beta$-decay process is described in terms of vector, and axial-vector couplings, with the known (V-A) structure generating the observed parity violation. There is, however, no a priori reason for only these terms to be present, and indeed, certain theoretical models introduce additional, Lorentz-Invariance conserving, couplings in the form of scalar, tensor, and pseudo-scalar contributions. The existence of such terms, would be immediate indication of beyond SM physics, and constraining them places bounds on the energy scale of such new physics~\cite{2019arXiv190702164C,GONZALEZALONSO2019165}.

At the low momentum transfers of $\beta$-decay (in the SM or any extension based on the exchange of massive bosons), the $\beta$-decay Hamiltonian can be expressed as a sum of possible contact 
interactions:
\begin{equation}
\CH_{\rm{int}}=\sum_X\l\bar{\psi}_p O_X \psi_n\r\l C_X\bar{\psi}_e O_X\psi_\nu
+C'_X \bar{\psi}_e O_X \gamma^5 \psi_\nu\r,
\end{equation}
where $O_X$ denotes operators with the different possible Lorenz transformation properties: 
vector (V), axial vector (A), tensor (T), scalar (S), and pseudoscalar (P), and the $\psi$s denote the lepton and nucleon fields. 
This $\beta$-decay formalism and its consequences for Standard Model tests is discussed thoroughly in Refs.~
\cite{Severijns:2013iza,Severijns:2011zz,Severijns:2006dr,Herczeg,Lee:1956qn}.   In the SM, the interaction between quarks and leptons is “V-A”
, meaning $C_V =  - C'_V = 1$ and $C_A= -C'_A \approx -1.27$, with $C_S$ and $C_T$ terms equal to zero, and emitting only left-handed 
neutrinos in $\beta$-decay.  

The most general expression for the nuclear $\beta$-decay rate, $\Gamma$, in terms
of the angular orientations and distributions of the leptons, 
for unpolarized nuclei, is given by~\cite{Jackson:1957zz}:
%%%%
\begin{equation}
\label{eq:betarate}
\Gamma dE_e d\Omega_e d\Omega_\nu\propto\left[1+a_{\beta\nu} \frac{\vec{p}_e\cdot\vec{p}_\nu}{E_e E_\nu}+b_F\frac{m_e}{E_e}\right],
\end{equation}
where $\vec{p}$($E$) are the lepton momenta (energies).

\begin{figure}[ht]
    \centering
    \includegraphics[width=0.75\textwidth]{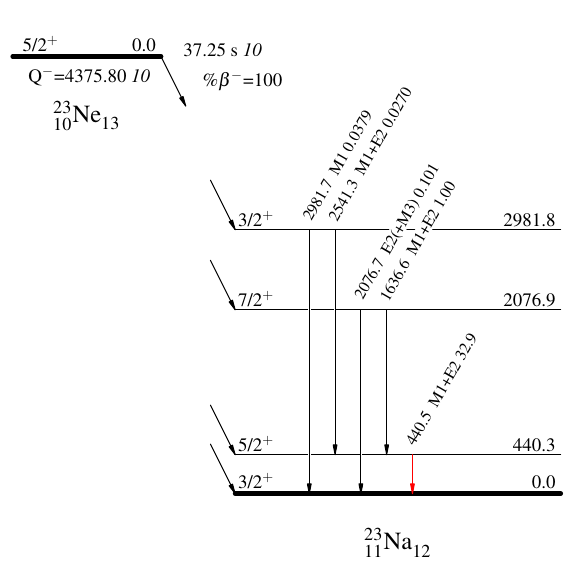}
    \caption{The $^{23}$Ne decay scheme, adapted from \cite{2021-Shamasuzzoha}.}
    \label{fig:23NeDecay}
\end{figure}

The correlation coefficients are given in~\cite{Jackson:1957zz}; e.g., $a_{\beta\nu}$,
the electron-neutrino correlation coefficient is 

\begin{eqnarray} \label{eq:a}
\nn a_{\beta\nu}&=&\left[\left|M_F\right|^2\l\left|C_V\right|^2+\left|C'_V\right|^2-\left|C_S\right|^2-\left|C'_S\right|^2\r\right.\\
&&\left.-\frac{1}{3}\left|M_{GT}\right|^2\l\left|C_A\right|^2+\left|C'_A\right|^2-\left|C_T\right|^2-\left|C'_T\right|^2\r\right]\xi^{-1},
\end{eqnarray}
$M_F$ and $M_{GT}$ are Fermi and Gamow-Teller (GT) matrix elements, and $\xi$ is a normalization coefficient. Note that Eq.~(\ref{eq:a}) shows the bare coupling, in 
the actual measurement additional effects (such as electromagnetic corrections, and recoil order corrections, stemming from the finite mass of the nucleus) modify 
the correlation and must be considered (see below).
As shown in~\cite{PhysRevC.94.035503} a measurement of the recoil ion spectra in $\beta$-beta decay encodes information about both the Fierz interference term ($b_F$), and the $\beta$-neutrino correlation (a$_{\beta\nu}$), 
the correlation between the directions of the outgoing $\beta$ particle and neutrino. This is
in contrast to the measurement of the $\beta$ particle spectra, where $a_{\beta\nu}$ is integrated out. Thus, a measurement of the recoil ion spectra is a very sensitive physics probe
of the interaction. A problem inherent in many such measurements arises, since some nuclei undergo $\beta$-decay to excited states of the daughter nucleus. In such processes, one must have excellent knowledge of the relevant branching ratios to properly interpret the data, since the 
particle spectra are different between the different branches. One such highly precise measurement of the recoil ion spectra is the decay of $^{23}$Ne performed by Carlson {\it et al.}~\cite{PhysRev.132.2239}, but unfortunately, the uncertainty in the branching ratio (BR) to the first excited state of the daughter $^{23}$Na, limited the extraction of $a_{\beta\nu}$ to a relative uncertainty of approximately 10\%, which is only sufficient for rudimentary 
constraints on the physics involved. Modern extractions of the couplings require uncertainties of order $\mathcal{O}$(10$^{-3}$) to be competitive. Figure~\ref{fig:23NeDecay}, 
adapted from~\cite{2021-Shamasuzzoha} shows the decay scheme of $^{23}$Ne to $^{23}$Na. 

In anticipation of new results from a measurement which is underway~\cite{ohayon2018weak, PhysRevC.101.035501} and to improve on the existing results from Carlson, we remeasured the BR of the $\beta$-decay to the first excited state of $^{23}$Na from the $\beta$-decay of $^{23}$Ne.

In addition, a new theory prescription \cite{glickmagid2021formalism} was used to calculate the recoil order corrections. As $\beta$-decays are characterized by a small momentum transfer $q$, we can identify a hierarchy of small parameters associated with $\beta$-decay observables, such as $\epsilon_{qr}\equiv qR$ and $\epsilon_{\text{recoil}}\equiv q/m_{N}$ (with R the nuclear radius, and $m_N$ the nucleon mass). According to that formalism, the Gamow-Teller (GT) general $\beta$-decay rate from Eq.~\eqref{eq:betarate}, should be rewritten as:
%%%
\be
\label{eq:corrected_betarate}
\Gamma dE_e d\Omega_e d\Omega_\nu&\propto&
\left(1+\delta_{1}\right)\left[1+a_{\beta\nu}\left(1+\tilde{\delta}_{a}\right)\frac{\vec{p}_e\cdot\vec{p}_\nu}{E_e E_\nu}+\left(b_F+\delta_{b}\right)\frac{m_e}{E_e}\right]\text{,}
\ee
where the different $\delta$s are shape and recoil next-to-leading-order (NLO) corrections, associated with the small parameters.
These $\delta$ corrections are influenced by the $\beta$-particle energy $E_e$, a factor that was overlooked by previous corrections used to analyze experiments~\cite{Holstein:1974zf}, and calculating them requires precise calculations of higher order matrix elements.
For example, for the ground state of $^{23}$Na, we obtained $|\delta_{1}| \lesssim 0.05$ , $|\tilde{\delta}_{a}| \lesssim 0.1$, and $|\delta_{b}|\approx 0.006$. This nuclear theory small parameter approach also allows us to determine the uncertainty of the corrections, which turns out to be about $0.0004$. This robust assignment of theoretical uncertainties is crucial to make improvements in $a_{\beta\nu}$ (full details of this method are described in the Methods section below).

\begin{figure}[ht] 
    \centering
    \includegraphics[width = 0.52\textwidth]{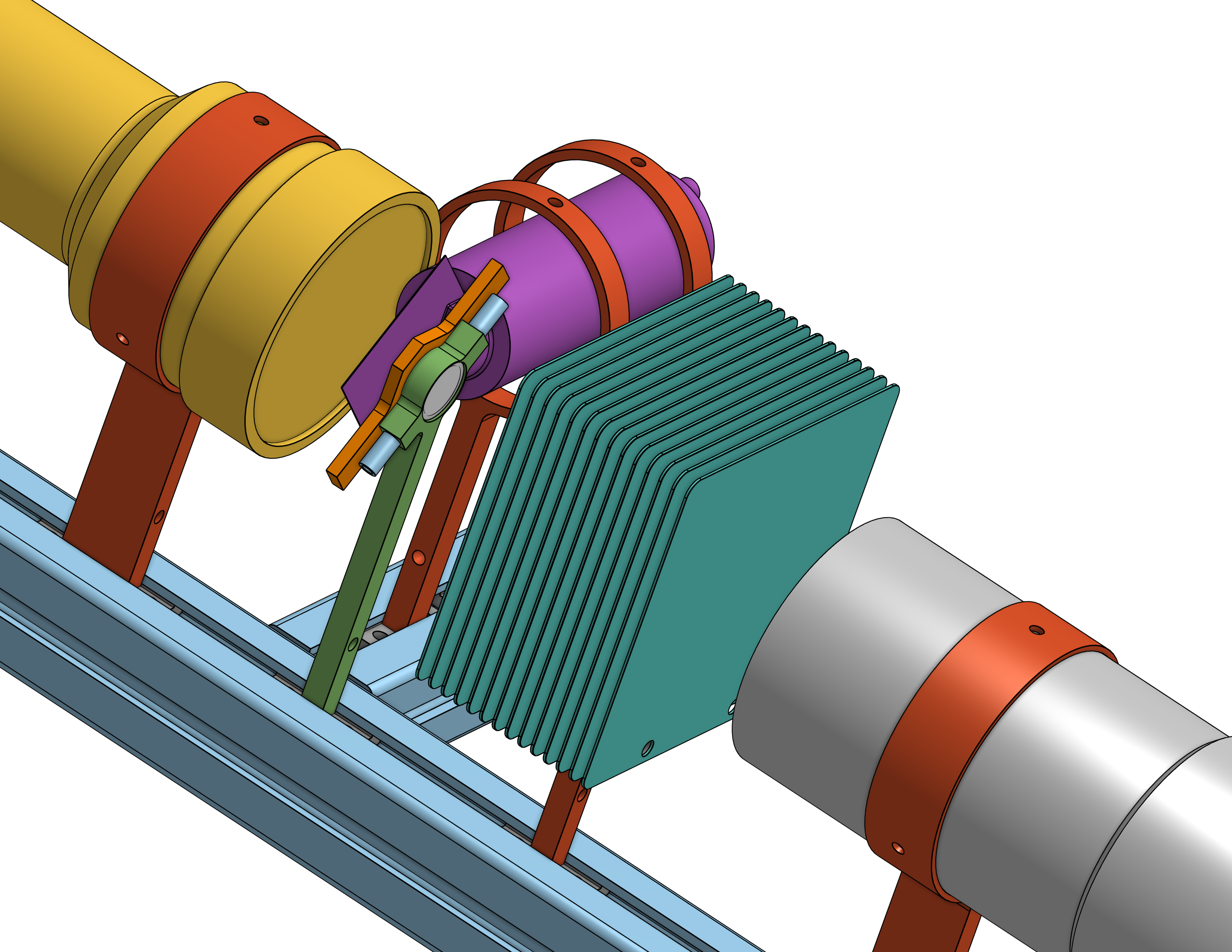}

    \caption{A schematic of the experimental system. To the top left (magenta and gold) are the scintillators used to tag the $\beta$ particles, to the bottom right (gray) is the HPGe detector, the teal foils are aluminum sheets used to block the $\beta$ particles from hitting the HPGe. The measurement cell (green) and a collimator (orange) can be seen in the middle of the figure. }
    \label{Fig:MeasCell}
\end{figure}

The $^{23}$Ne ($\tau_{1/2}$ = 37.24 s) atoms were produced via an (n,p) reaction on a natural abundance NaCl target. The neutrons were produced via a (d,n) reaction on a liquid lithium target “LiLiT”~\cite{2013-Halfon,2014-Halfon,2015-Halfon,2015-Paul,2019EPJALiLiT}. The neon atoms diffused from the target and were transported via a vacuum line backed by a turbomolecular pump whose output was connected to the measurement cell. Fig. \ref{Fig:MeasCell} shows a schematic of the measurement setup.
The measurement cell had a diameter of 17 mm and a thickness of 6 mm, and was 
made of aluminum with thin (75 $\mu$m) Be windows. Two plastic scintillators 
were installed on one side, which detected the $\beta$ particles. The plastic scintillators consisted of a thin (0.5 mm, EJ-212) and a thick (2 cm, EJ-200) scintillator, each read out by a photomultiplier tube. Data was recorded when any detector registered an event above threshold, and a $\beta$ event was defined as a hit in both detectors within a 1 microsecond coincidence window. On the other side of the measurement cell, a High Purity Germanium detector (Ortec 4083 HPGe) was used to detect $\gamma$ rays in coincidence with the emitted $\beta$ particles, whenever an event occurred from the excited state decay (See the supplemental information for a full description).  The branching ratio is defined as the ratio of the number of events where a $\gamma$-particle, of the appropriate energy (i.e., 440 keV), is detected in coincidence with a $\beta$ particle (N$_{\beta\gamma}$) over the number of detected $\beta$ particles (N$_\beta$), normalized to the HPGe detector efficiency (i.e., 440 keV), see the supplemental information for more details;
\be
    BR_{440 keV}=\frac{{\rm N}_{\beta\gamma}}{\epsilon_{\gamma} {\rm N}_{\beta}}
\ee

\begin{figure}[ht]
\centering
   \includegraphics[width= 0.9\textwidth]{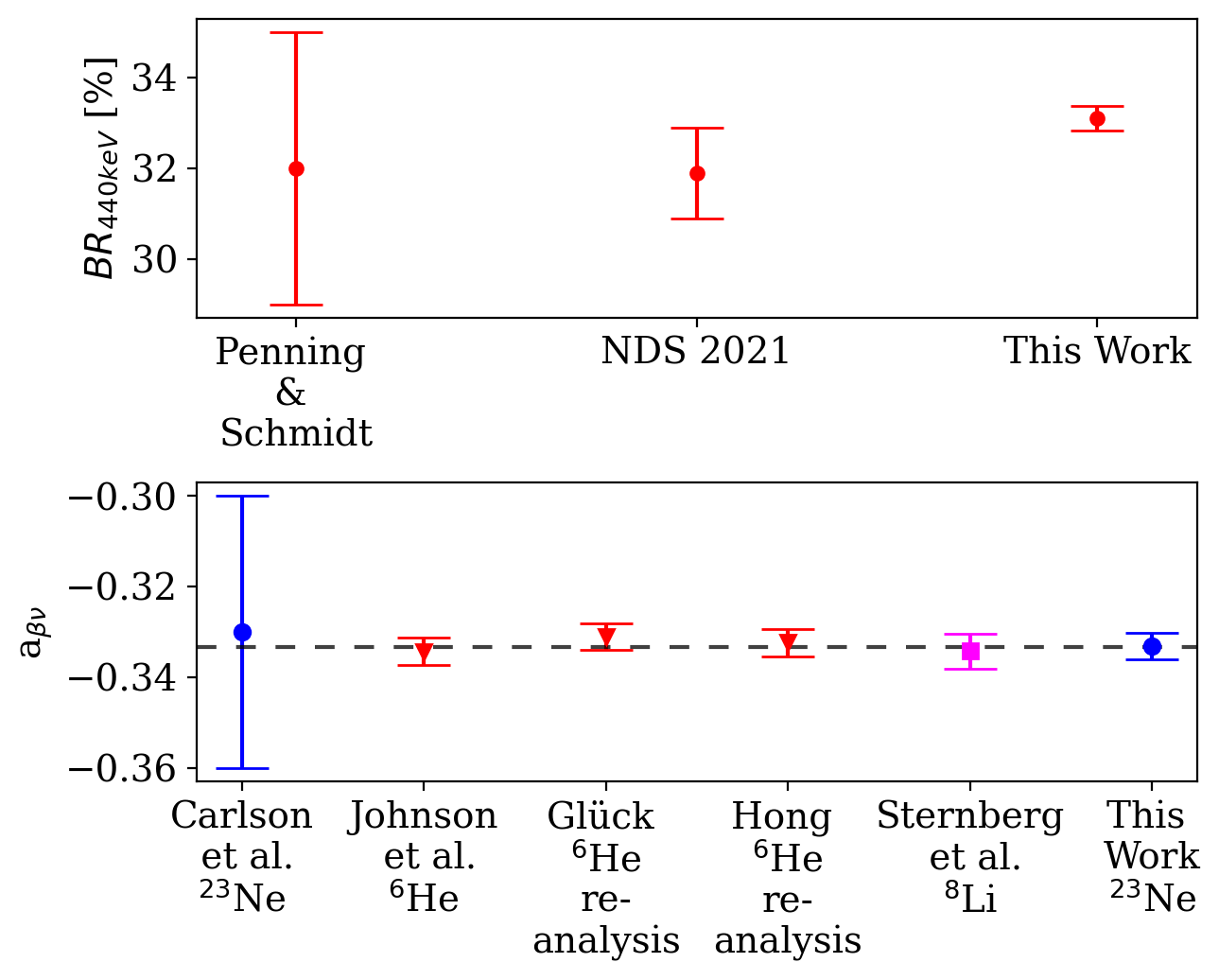}%
    \caption{ 
    Top: $^{23}$Ne Branching ratio measurements (\cite{PhysRev.105.647}, \cite{2021-Shamasuzzoha}, and this work). Bottom: a$_{\beta\nu}$ extractions (\cite{PhysRev.132.2239}, \cite{1963-Johnson}, \cite{GLUCK1998493},
    \cite{Hong:2016qff},
    \cite{PhysRevLett.115.182501}, and this work), the dashed line
    is the SM value.}
    \label{Fig:Hist}

\end{figure}

The BR$_{440keV}$ obtained at two different distances for the HPGe detector were treated as independent measurements. We obtained a BR$_{440keV}$ of 0.3293 $\pm$ 0.0027 at 16 cm and a BR$_{440keV}$ of 0.3354 $\pm$ 0.0043 at 20 cm. We take the weighted average of these two results (where the uncertainty is inflated
by $\sqrt{\chi^2_\nu}$) and obtain 
\be
BR_{440 keV}=0.3310 \pm 0.0027.
\ee
This result is an improvement by factor 3 over the currently 
accepted value~\cite{2021-Shamasuzzoha} which relies on assuming 
the correctness of the SM, and an order of magnitude improvement 
over the best direct measurement~\cite{PhysRev.105.647}. See Fig.~\ref{Fig:Hist} (Top) for a comparison of the various extractions. The measured branching ratio was used to determine the $\beta$ intensities following~\cite{2021-Shamasuzzoha}, see the supplemental information for an update of the 
$^{23}$Ne $\beta$ intensities.

Using the newly  determined $\beta$ intensities, we now reanalyze the data from~\cite{PhysRev.132.2239} (where recoil ions from the decay 
of $^{23}$Ne were analyzed for charge and energy using a combination of electrostatic and magnetic spectrometers, see~\cite{1963-Johnson,PhysRev.132.2239} for complete details) to obtain an updated value for $a_{\beta\nu}$.
We extract the value of $a_{\beta\nu}$ for $^{23}$Ne by performing a Bayesian analysis (using PyStan~\cite{pystan}). 
We generate spectral shape templates for the decay to the individual states of $^{23}$Na, for Gamow-Teller (a$_{\beta\nu}$=-1/3),Tensor (a$_{\beta\nu}$=1/3) decays, and a template containing only the Fierz term,  using a Monte-Carlo simulation which includes hard and soft electromagnetic 
corrections~\cite{GLUCK1997223, CALAPRICE1976301}, recoil from the excited state $\gamma$ particles, with the Fermi function from~\cite{Schopper:1969jkp}, and new theoretical calculations for
the recoil order corrections (see Methods).

\begin{figure}[ht]
    \subfloat[]{\includegraphics[width=0.44\textwidth]{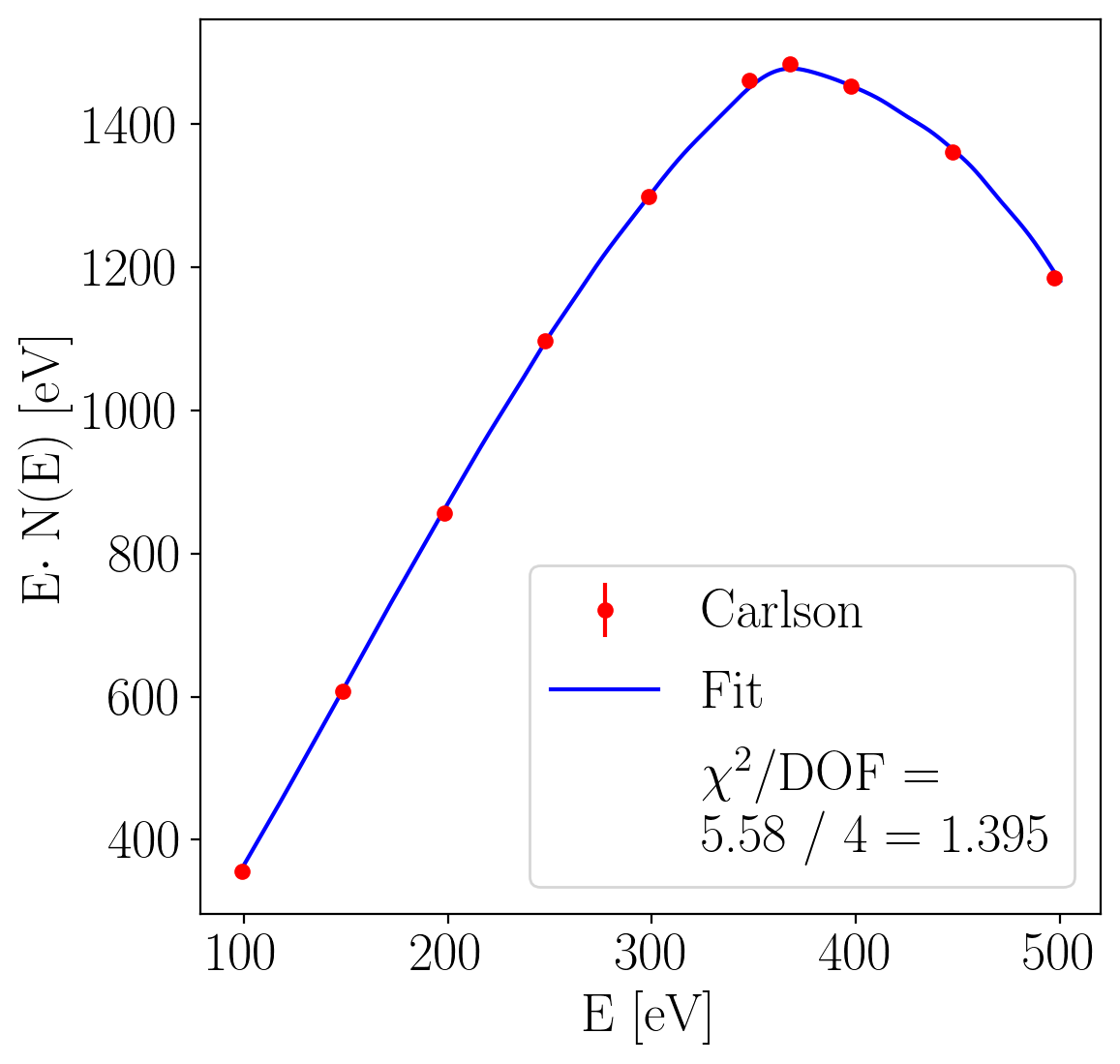}}%
    \qquad
    \subfloat[]{\includegraphics[width=0.45\textwidth]{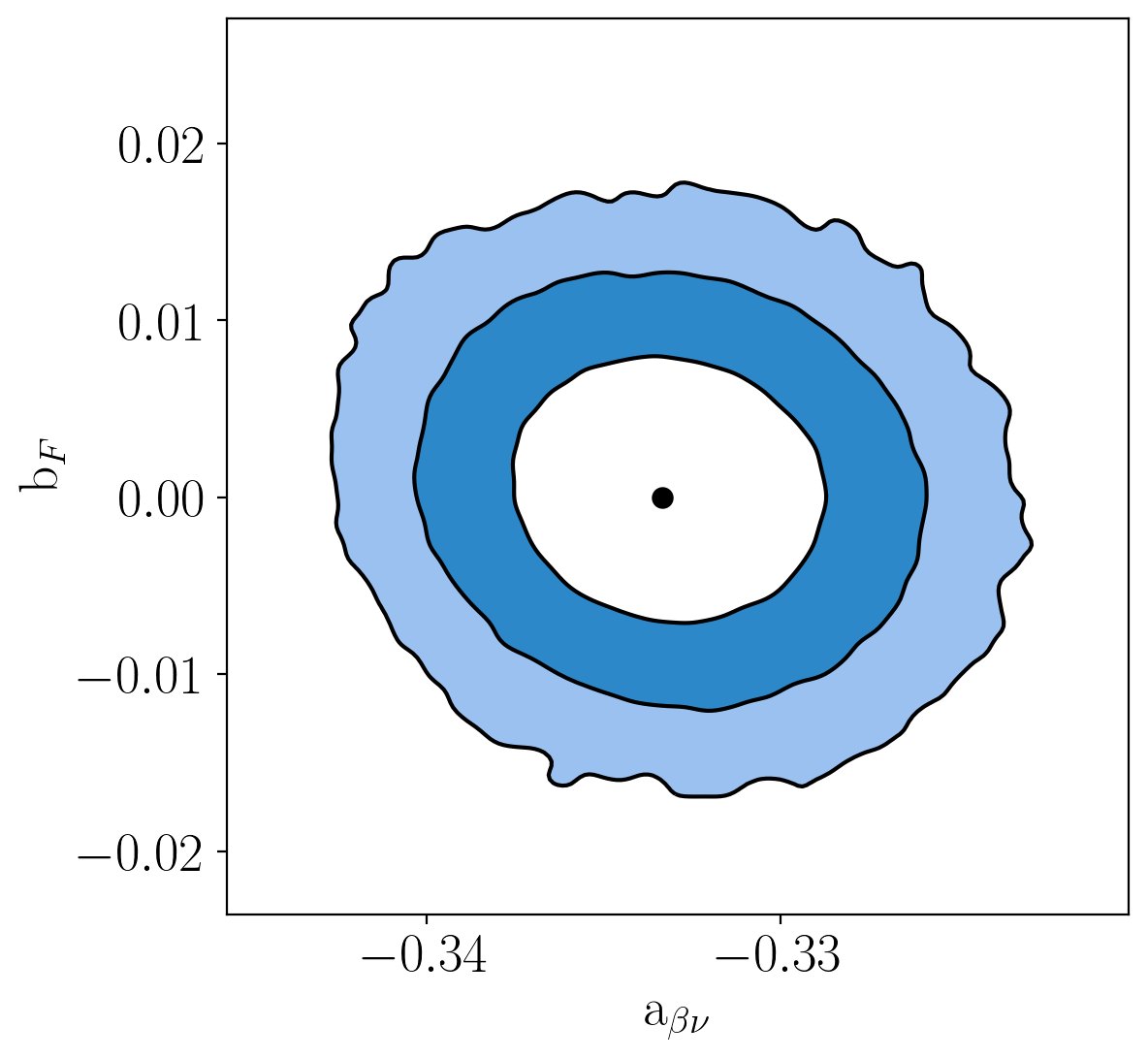}}%
    \caption{Results of the fit to the data from~\cite{PhysRev.132.2239}. (a) The fit. (b) Joint PDF for a$_{\beta\nu}$ and b$_F$, the black dot indicates the SM values.}
    \label{23Ne_Fit}
%    \vspace{-0.8cm}
\end{figure}

We fit the Carlson data~\cite{PhysRev.132.2239} to the form:
\begin{gather}
\nn E\cdot N(E)=E\cdot\eta\sum_{i=0}^3 BR_i \left[\lambda N_{i}^{GT}(E)+\right.\\
\left.(1-\lambda)N_{i}^{T}(E)+b_F N_{i}^{\rm Fierz}(E)\right],\\
BR_0\equiv 1-BR_1-BR_2-BR_3,
\end{gather}
where E is the recoil kinetic energy, $N_{0}$, $N_1$, $N_2$, $N_3$, are the spectral shapes 
for the decay to the ground, 1st, 2nd, and 3rd excited states, respectively, with the 
superscript indicating a Gamow-Teller (axial), Tensor, or Fierz term only (no isotropic or correlation terms) decay (for the axial and tensor terms we set b$_F$=0), the $BR$s are
the branching ratios, $\eta$ is a normalization constant, $\left[-1/3\cdot\lambda+1/3\cdot(1-\lambda)\right]$ is 
the $\beta$-neutrino correlation (a$_{\beta\nu}$), and $b_F$ is the Fierz term. 
The measured branching 
ratios are used as normally distributed priors in our fit procedure. Figure~\ref{23Ne_Fit} shows the results of the fit, contour plot for the joint probability density function (PDF) of a$_{\beta\nu}$ and b$_F$, the contours show the 68.3\%, 95.5\%, and 99.7\% confidence regions.

\renewcommand\floatpagefraction{.1}%

We obtain the values:
\begin{eqnarray}
a_{\beta\nu}&=&-0.3331\pm0.0028\pm 0.0004\pm0.0002,\\
b_F&=&0.0007\pm0.0049\pm 0.0003\pm0.0001,
\end{eqnarray}
where the second uncertainty is related to the potential effects of pressure on the data in the Carlson experiment (see~\cite{PhysRev.132.2239}), and the third uncertainty is the theory uncertainty for the 
higher order corrections (see Methods section and the supplemental information), we note that the uncertainty on b$_F$ is a factor of 2 better than the current 
best available measurement in a pure GT decay~\cite{20FThesis}, and 
is comparable to the best extractions of a$_{\beta\nu}$ in such a decay~\cite{PhysRevLett.115.182501,GLUCK1998493,Hong:2016qff,PhysRevLett.115.182501}, see Fig~\ref{Fig:Hist} (Bottom).

Table~\ref{tab:err_budget} details the error budget for the measurement of the branching 
ratio and for the extraction of the $\beta$-neutrino correlation. See the supplemental 
information for more details on the individual uncertainties. 

\begin{table}[ht]
    \centering
    \begin{tabular}{l l c}
    & Parameter & Relative uncertainty (\%)\\ 
        \toprule
    {\bf BR$_{440keV}$ - 16 cm}&        
         440 keV events ($N_{\beta\gamma}$)& 0.51 \\
    &     $\beta$ events ($N_{\beta})$& 0.02 \\
    &     Higher state feeding & 0.12 \\
    &     $\beta$ threshold & 0.08\\
    &     Contaminants & 0.02 \\
    &     HPGe efficiency ($\varepsilon_\gamma$) & 0.6 \\
         \midrule
    &     Total & $0.81$ \\
            \midrule
    {\bf BR$_{440keV}$ - 20 cm}&        
         440 keV events ($N_{\beta\gamma})$& 0.81 \\
    &     $\beta$ events ($N_{\beta})$& 0.02 \\
    &     Higher state feeding & 0.12 \\
    &     $\beta$ threshold & 0.08\\
    &     Contaminants & 0.02 \\
    &     HPGe efficiency ($\varepsilon_\gamma$) & 0.99 \\
         \midrule
    &     Total & $1.29$ \\
            \midrule
    {\bf a$_{\beta\nu}$}&        Fitting + Branching ratio & 0.84 \\
    &        Q Value & 0.01 \\
    &        Pressure effects & 0.12 \\
    &        Recoil order corrections & 0.06\\
            \midrule
    &        Total & 0.85 \\
            
        \bottomrule
    \end{tabular}
    \caption{Error budget for the branching ratio measurement and for the extraction of a$_{\beta\nu}$.}
    \label{tab:err_budget}
\end{table}

Figure~\ref{Fig:Hist} compares the results from this work to previous measurements 
of the branching ratio (Top) and to extractions of  a$_{\beta\nu}$ in  
pure GT decay~\cite{1963-Johnson,PhysRev.132.2239,PhysRevLett.115.182501}, and two recent reanalyses of the $^{6}$He dataset which consider additional
corrections~\cite{GLUCK1998493,Hong:2016qff} (Bottom), 
the improvement in the measurements, as well as the consistency with 
the best world data, are evident. 

\begin{figure}[ht]
\centering
   \includegraphics[width= 0.75\textwidth]{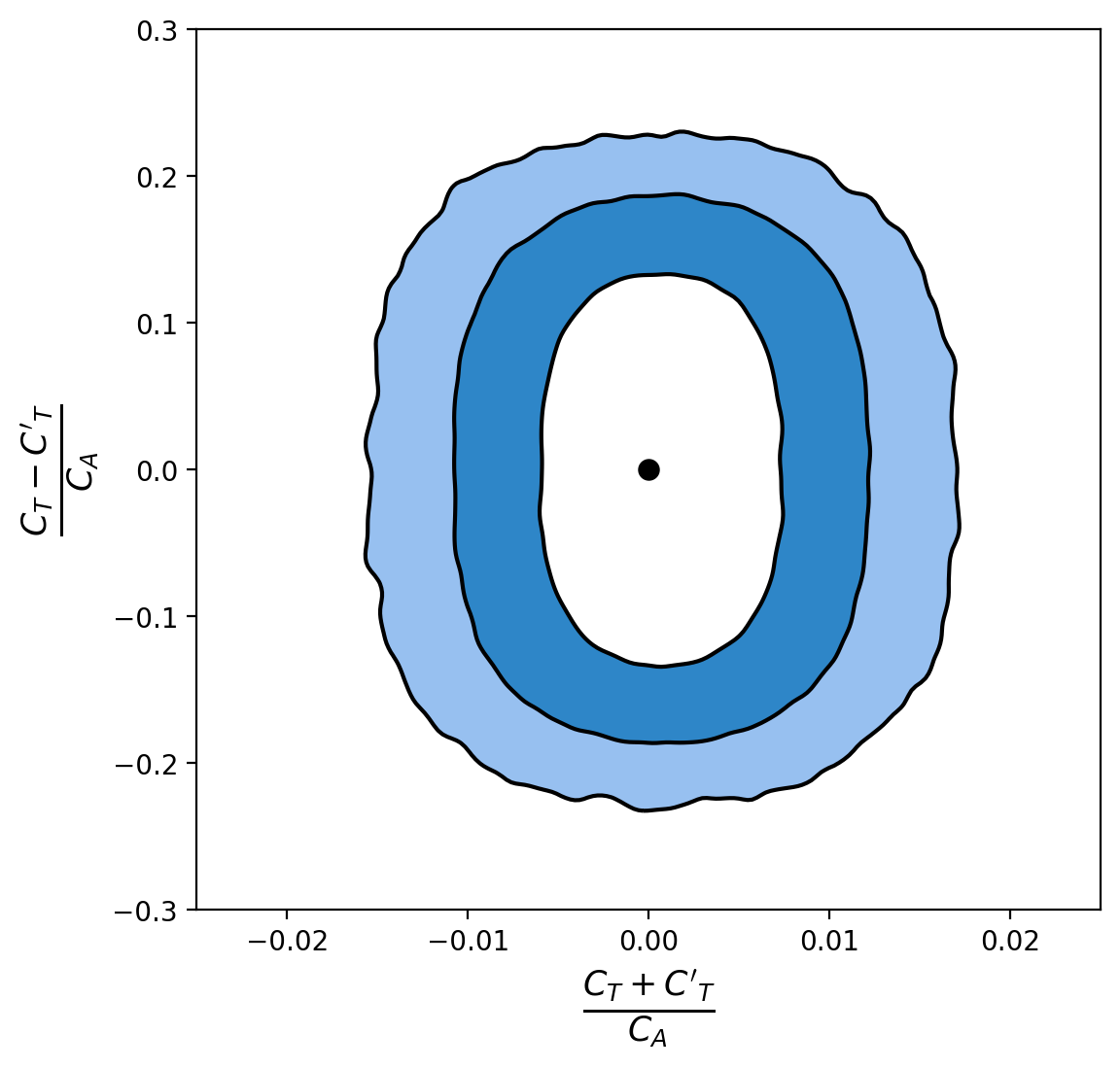}%
    \caption{ 
    Extraction of right- and left-handed combinations of the tensor couplings $C_T$, $C'_T$. The black dot indicates Standard Model values ($C_T=C'_T=0$)}
    \label{Fig:CT_CTp}
%    \vspace{-0.8cm}
\end{figure}

Using the extracted values for a$_{\beta\nu}$ and b$_F$, and assuming 
time-reversal invariance (${\cal{I}}m(C_i)={\cal{I}}m(C'_i)=0$) and maximal parity violation for the axial interaction ($C_A=C'_A$), 
one can write following~\cite{Jackson:1957zz,GONZALEZALONSO2019165}:
\begin{gather}
a_{\beta\nu}=-\frac{1}{3}\left(\frac{2-\frac{|C_T|^2+|C'_T|^2}{C_A^2}}{2+\frac{|C_T|^2+|C'_T|^2}{C_A^2}}\right)\\
b_F=\sqrt{1-\alpha^2 Z^2}\frac{C_T+C'_T}{C_A},
\end{gather}
where $\alpha$ is the fine structure constant, and Z is the charge of the daughter 
nucleus (Z=11 for $^{23}$Na). 
Performing a Bayesian fit (using uninformative, broad Gaussian, priors) it is possible to place constraints on 
the tensor couplings $C_T/C_A$, $C'_T/C_A$, and their combinations which embody interaction involving left or right-handed neutrinos. Figure~\ref{Fig:CT_CTp} shows the 68.3\%, 95.5\%, and 99.7\% confidence regions for this extraction, which yields:
\begin{gather}
%\frac{C_T}{C_A}=0.000192\pm0.044540\\
%\frac{C'_T}{C_A}=0.000178\pm0.044496\\
\frac{C_T+C'_T}{C_A}=0.0007\pm0.0049\\
\frac{C_T-C'_T}{C_A}=0.0001\pm0.0823
\end{gather}

For a recent review of the world data 
for the various coefficients see~\cite{GONZALEZALONSO2019165,Falkowski:2020pma}, note that the new result compares favorably to the uncertainties achieved in global 
fits which take into account results from multiple experiments~\cite{Falkowski:2020pma}).

In conclusion, a new measurement of the $^{23}$Ne $\beta$-decay branching ratio 
to the first excited state was performed, improving on the known value by {\it an order of magnitude}. A combination of this new result with  novel theoretical calculations of the recoil order correction and existing high-precision measurements of the recoil ion spectra from $^{23}$Ne have conclusively demonstrated the applicability of this method to extract angular correlations in $\beta$-decay to extremely high-precision. Such high-precision measurements are required to probe new physics at energy scales still not excluded by existing measurements. While our results are consistent with the Standard Model of particle physics, new measurements are underway, which are
anticipated to improve on this result by yet another order of magnitude~\cite{PhysRevC.101.035501}. 

\putbib[isotrap]
\end{bibunit}

\begin{bibunit}
\section*{Methods}

\subsection*{$^{23}$Ne production}
As mentioned, $^{23}$Ne was produced via the (n,p) reaction on 
the $^{23}$Na content in NaCl~\cite{mardor2018soreq, MISHNAYOT2020164365}. See supplemental information for complete
details.

\subsection*{$^{23}$Ne BR measurement}

To reduce systematic effects, the measurements were taken with the HPGe detector at two distances relative to the measurement cell (16 cm, and 20 cm). For each of those distances, the HPGe detector efficiency was fit for at the two relevant energies (440 and 1636 keV) using a set of calibrated sources (calibrated to between 1\% and 2.5\%). See the supplemental information for more details.

\subsection*{Recoil order calculations and accuracy estimation}

We use a new theory formalism~\cite{glickmagid2021formalism} to calculate the recoil order corrections and to assess the accuracy of the nuclear-structure weak interaction effects. According to that formalism, the shape and recoil NLO corrections presented in Eq.~\eqref{eq:corrected_betarate} are:
\begin{align}
\label{eq:delta}
\delta_1 & =
\frac{2}{3}
\left[-E_0
\frac{\left\langle \left\Vert \hat{C}_{1}^{A}/q\right\Vert \right\rangle }
{\left\langle \left\Vert \hat{L}_{1}^{A}\right\Vert \right\rangle }
+\sqrt{2}\left(E_0-2E_e\right)
\frac{\left\langle \left\Vert \hat{M}_{1}^{V}/q\right\Vert \right\rangle }
{\left\langle \left\Vert \hat{L}_{1}^{A}\right\Vert \right\rangle }
\right],
\nonumber \\
\tilde{\delta}_{a} & =
\frac{4}{3}
\left[2E_0
\frac{\left\langle \left\Vert \hat{C}_{1}^{A}/q\right\Vert \right\rangle }
{\left\langle \left\Vert \hat{L}_{1}^{A}\right\Vert \right\rangle }
+\sqrt{2}\left(E_0-2E_e\right)
\frac{\left\langle \left\Vert \hat{M}_{1}^{V}/q\right\Vert \right\rangle }
{\left\langle \left\Vert \hat{L}_{1}^{A}\right\Vert \right\rangle }
\right],
\nonumber \\
\delta_{b} & =
\frac{2}{3}m_{e}
\left[\frac{\left\langle \left\Vert \hat{C}_{1}^{A}/q\right\Vert \right\rangle }
{\left\langle \left\Vert \hat{L}_{1}^{A}\right\Vert \right\rangle }
+\sqrt{2}
\frac{\left\langle \left\Vert \hat{M}_{1}^{V}/q\right\Vert \right\rangle }
{\left\langle \left\Vert \hat{L}_{1}^{A}\right\Vert \right\rangle }
\right].
\end{align}
Here, $\left(E_0,\vec{q}\right)=\left(E_e,\vec{p}_e\right)+\left(E_{\nu},\vec{p}_{\nu}\right)$
is the 4-momentum transfer in the process, 
and
$\left\langle \left\Vert \hat{L}_{1}^{A}\right\Vert \right\rangle$,
$\left\langle \left\Vert \hat{C}_{1}^{A}/q\right\Vert \right\rangle$
and
$\left\langle \left\Vert \hat{M}_{1}^{V}/q\right\Vert \right\rangle$
are reduced matrix elements of multipole operators
between the initial and final nuclear wave functions (for explicit expressions of the operators see supplemental). These NLO corrections are caused by the multipole operators $\hat{C}_{1}^{A}$ and $\hat{M}_{1}^{V}$, which in the nomenclature of~\cite{glickmagid2021formalism}, are both dominated by two small dimensionless parameters, $\epsilon_{\text{recoil}}$ and $\epsilon_{qr}\epsilon_{\text{NR}}$.
Here $\epsilon_{\text{recoil}}\approx 0.005$ and $\epsilon_{qr}\approx 0.07$ for the ground state of $^{23}$Na (and smaller for higher excited states), and $\epsilon_{\text{NR}} \equiv P_{\text{Fermi}}/m_{N}\approx0.2$.
To these NLO recoil and shape corrections, we add NLO Coulomb recoil corrections, calculated explicitly from the approximation displayed in the appendix of Ref.~\cite{calaprice1976weak}, based on Holstein's formalism~\cite{holstein1974electromagnetic}.
Using the new formalism approach~\cite{glickmagid2021formalism}, along with Coulomb terms presented in Refs.~\cite{hayen2018high, hayen2020consistent}, we determine that for $^{23}$Ne, next-to-next-to-leading-order (NNLO) corrections would be of the order of $\frac{1}{15}E_0 R \alpha Z \approx 0.0004$ (with $\alpha$ the fine-structure constant), or smaller.

The wave functions and the matrix elements of the operators of the $^{23}$Ne $\beta$-decay to $^{23}$Na, have been calculated using USDB \cite{brown2006new} and USD \cite{wildenthal1984empirical} interactions, both gold-standard shell model interactions. Detailed information is provided in the supplemental.
Wave functions and matrix elements for $^6$He $\beta$-decay to $^6$Li, used to validate the analysis (see supplemental), were derived within the ab initio no-core shell model (NCSM)~\cite{BARRETT2013131} using $\chi$EFT interactions~\cite{N2LOopt,N2LOsat}. Detailed information, including accuracy estimation of the $^6$He calculations, can be found in~\cite{glickmagid2021nuclear}.

%\subsection*{$a_{\beta\nu}$ and $b_F$ extraction}
%\printbibliography
\putbib
\end{bibunit}

%\end{refsection}

\section*{Data availability}
The data presented in the figures of this Article are available from the corresponding authors upon reasonable request.
\section*{Code availability}
All custom code used to support claims and analyze data presented in this article is available from the corresponding authors upon reasonable request.

\section*{Author contributions}
G.R. and B.O. suggested the research. Y.M., H.R., B.O., G.R., J.H., N.S., A.G., S.V., T.H., L.W., H.D., M.B., A.K., and B.K. contributed to the experimental design, construction, and data collection. Y.M., H.R., G.R., and J.H. analyzed the data. D.Gazit, A.G-M., C.F., D.Gazda, P.G., J.M., and P.N. performed theoretical calculation and numerical simulations. N.S., A.G., G.R., and Y.M, wrote the simulation code used to analyze the results. All authors contributed
to the discussion of the results and the manuscript.

\section*{Competing interests}
The authors declare no competing interests.
\section*{Acknowledgments}
The work of Y. M., A.G-M., and G. R. has been supported by the Israeli Science Foundation under ISF grant 139/15, and the European 
Research Council (Grant No. 714118 TRAPLAB). G.R. and S.V. acknowledge support
from the Pazy Foundation. This work was performed under the auspices of the U.S. Department of Energy by Lawrence Livermore National Laboratory under Contract DE-AC52-07NA27344 and was supported by the LLNL-LDRD Program under Project No. 19-ERD-011. For part of his involvement with this work, B.O. was supported by the Ministry of Science and Technology, under the Eshkol Fellowship. A.G-M’s research was partially supported by the Ministry of Science and Technology, Israel, under the Navon scholarship. We wish to acknowledge the support of the ISF grant no. 1446/16 (D.Gazit and A.G-M.), the Swedish Research Council, Grant No. 2017-04234 (CF and D.Gazda), the Czech
  Science Foundation GA\v{C}R grant No.\ 19-19640S (D.Gazda), and the NSERC Grants No. SAPIN-2016-00033 (P.G. and P.N.) and
  PGSD3-535536-2019 (PG).
  TRIUMF receives federal funding via a contribution agreement with the National Research Council of Canada.  Computing support came from an INCITE Award on the Summit supercomputer of the Oak Ridge Leadership Computing Facility (OLCF) at ORNL, and from Westgrid and Compute Canada.
  Parts of the computations were enabled by resources provided by the Swedish National Infrastructure for Computing (SNIC) at Chalmers Centre for Computational Science and Engineering (C3SE), the National Supercomputer Centre (NSC) partially funded by the Swedish Research Council.
  CF acknowledges support from the Swedish Research Council, Grant No. 2017-04234.
  This work was supported by the Spanish MICINN through the “Ram\'on y Cajal” program with grant RYC-2017-22781, the AEI “Unit of Excellence Mar\'ia de Maeztu 2020-2023” award CEX2019-000918-M and the AEI grant FIS2017-87534-P.
The authors thank the SARAF technical crew and operators, the LiLiT 
developers and operators, and the SNRC technical and workshop staff.

\end{document}